\documentclass[reprint,superscriptaddress]{revtex4-1} 
\usepackage{amsmath,amssymb,graphicx,hyperref}
\newcommand{\unit}[1]{\ensuremath{\, \mathrm{#1}}}
\begin{document}

\title{Vortex-phase filtering technique for extracting spatial information \\from unresolved sources}

\author{Garreth J. Ruane}
\affiliation{Chester F. Carlson Center for Imaging Science, Rochester Institute of Technology, 54 Lomb Memorial Drive, Rochester, NY 14623, USA}
\affiliation{New York State Center for Complex Light, NY, USA}
\author{Prachyathit Kanburapa}
\affiliation{Chester F. Carlson Center for Imaging Science, Rochester Institute of Technology, 54 Lomb Memorial Drive, Rochester, NY 14623, USA}
\author{Jiaxuan Han}
\affiliation{Chester F. Carlson Center for Imaging Science, Rochester Institute of Technology, 54 Lomb Memorial Drive, Rochester, NY 14623, USA}
\author{Grover A. Swartzlander, Jr.}\email{Corresponding author: grover.swartzlander@gmail.com}
\affiliation{Chester F. Carlson Center for Imaging Science, Rochester Institute of Technology, 54 Lomb Memorial Drive, Rochester, NY 14623, USA}
\affiliation{New York State Center for Complex Light, NY, USA}

\begin{abstract}A white light vortex coronagraph was used to experimentally achieve sub-resolution detection.  The angular location of the centroid, $\gamma$, and the angular extent of circular pinhole sources, $\Theta$, were measured to within errors of $\delta\gamma = \pm0.015\,\lambda/D$ and $\delta\Theta = \pm0.026\,\lambda/D$.  This technique has two advantages over conventional imaging: simple power measurements are made and shorter exposure times may be required to achieve a sufficient signal-to-noise ratio.    
\end{abstract}


\maketitle 

This paper was published in Applied Optics and is made available as an electronic reprint with the permission of OSA. The paper can be found at the following URL on the OSA website: \url{http://www.opticsinfobase.org/ao/abstract.cfm?URI=ao-53-20-4503}. Systematic or multiple reproduction or distribution to multiple locations via electronic or other means is prohibited and is subject to penalties under law.

\section{Introduction}
The spatial resolution of an imaging system with a finite aperture is fundamentally limited by the wave nature of light.  However, optical pre-detection processing techniques may be employed to circumvent conventional resolution criteria, including magnitude and/or phase filtering \cite{Osterberg1949,Toraldo1952}.  Moreover, non-imaging sensors as described here may be advantageous for obtaining specific source information that is otherwise unresolved.  We experimentally demonstrate a vortex-phase filtering technique for extracting the centroid and spatial extent of an unresolved source distribution \cite{Swartzlander2011,Han2012thesis}.  Spatially integrated power measurements provide sub-resolution structure information akin to radial moments of the source (e.g. the radial variance).  This approach may be suitable for characterizing unresolved targets, such as near earth objects or small satellites. 

Applications of vortex phase elements span many areas of imaging science including optical spatial filtering \cite{Khonina1992}, phase contrast microscopy \cite{Furhapter2005}, and high-contrast astronomical imaging \cite{Mawet2005,Foo2005}.  The latter makes use of an optical vortex coronagraph (OVC).  The OVC is a Fourier filtering instrument wherein a focal plane vortex phase element acts to spatially separate light originating from a distant point source that coincides with the optical axis from nearby off-axis sources.  Consequently, the OVC is adept at small-angle high-contrast astronomical observations of stars \cite{Swartzlander2008} and exoplanets \cite{Serabyn2010}.  

The optical design demonstrated here is adopted from a recently introduced broadband scalar OVC \cite{Kanburapa2012conf,Kanburapa2012thesis,Errmann2013}.  This design incorporates a computer generated vortex hologram and a complementary dispersion compensating diffraction grating to form a vortex-phase filtering instrument that may be used with white light \cite{Leach2003,Mariyenko2005}.  We demonstrate using the broadband scalar OVC to determine the spatial extent of unresolved sources in the laboratory.  This technique leverages the remarkable sensitivity of the OVC to the angular position of point sources near the optical axis of the system.  

\section{Broadband vortex-phase filtering}
The optical layout of an OVC is a $4f$ lens system [see Fig. 1].  Lens L1, located at the entrance aperture (AP), focuses light onto a vortex phase element with transmission $t_m = \exp(im\theta')$, where $m$ is an integer known as topological charge and $\theta'$ is the azimuth in the $x', y'$ plane [see Fig. 1(b)].  For nonzero even values of $m$, all irradiance from a point source whose focal spot is centered on the phase singularity diffracts outside of the geometric image of AP.  The amplitude of the field at the exit pupil plane may be written in polar coordinates as 
\begin{equation}
\left|U_m\left(r'',\theta''\right)\right|=\left\{ \begin{matrix}
   \,\frac{R}{r''}\,\left|Z_{n}^{1}(\frac{R}{r''})\right|, & r''>R  \\
   0, & r''<R  \\
\end{matrix} \right.,
\end{equation}
where $m$ is an even nonzero integer, $R$ is the radius of AP, $n = |m|-1$, and $Z_{n}^{1}(R/r'')$ are the radial Zernike polynomials \cite{Carlotti2009}.  For $m = 2$, the exterior field amplitude is $|U_2\left(r'',\theta''\right)|={{\left(R/r''\right)}^{2}}$ [see Fig. 1(c)].  On the other hand, the focal spot due to an off-axis source is displaced away from the phase singularity at the origin in the $x', y'$ plane.  Consequently, the vortex phase element has little effect on the image of AP [see Fig. 1(d)].  By placing a second aperture at the $x'', y''$ plane, known in coronagraphy as the Lyot stop (LS), light originating from the on-axis source is obstructed, while light from off-axis sources is transmitted.  The radius of the LS must be less than or equal to the radius of AP to reject all of the light from an on-axis source.  Here $R_{LS}=R/2$ is chosen to remedy optical aberration effects.  Fig. 2 shows the normalized power transmitted through the LS from a distant point source, $P_m$, displaced from the optical axis by angle $\alpha$.  In the $m = 2$ case, the source is attenuated to $\sim50\%$ of the maximum value ($\sim0.25$) when $\alpha = \lambda/D$, where $\lambda$ is the wavelength and $D = 2R$.  Meanwhile, $P_m$ falls off asymptotically as $\alpha$ approaches zero.  Systems with higher $m$ values reject more light from off-axis sources \cite{Mawet2005,Foo2005}.

\begin{figure}[t!]
\centerline{\includegraphics[width=\columnwidth]{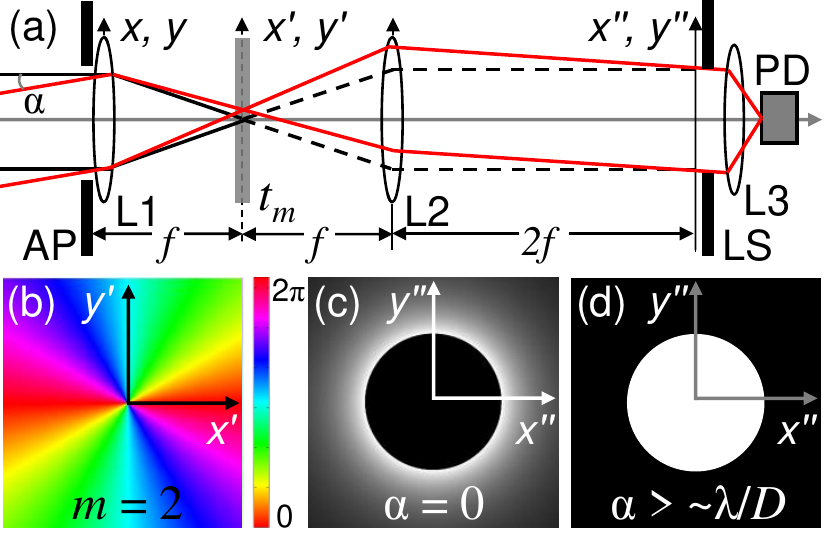}}
\caption{(a) Schematic of an OVC with lens L1 located at entrance aperture AP, vortex phase element with transmission $t_m = \exp(im\theta')$, and field lens L2.  The geometric exit pupil is located at the $x'', y''$ plane, where the Lyot stop (LS) truncates the field.  The spatially integrated power transmitted through the LS is recorded by a photo detector (PD).  (b) Phase profile of vortex phase mask with transmission $\exp(i2\theta')$.  (c) On-axis coherent light ($\alpha = 0$) is diffracted outside of the geometric exit pupil ($m = 2$ case shown).  (d) Tilted plane waves at the aperture form an approximate image of the entrance pupil at the $x'', y''$ plane.}
\end{figure}

The application described here takes advantage of the OVC's steep response to source displacement and therefore requires high precision vortex elements such as a spiral phase plate produced via electron beam lithography \cite{Swartzlander2008}, subwavelength diffraction gratings \cite{Mawet2005,Bomzon2001,Delacroix2013}, liquid crystal elements \cite{Marrucci2006,Mawet2009}, or photonic crystal elements \cite{Murakami2013}.  The spiral phase plate operates at a single wavelength, which may not be practical for white light systems.  The other approaches have reported spectral bandwidths as large as $\Delta\lambda/\lambda = 20\%$.  Here we employ a vortex hologram (VH) and dispersion compensation.  The light transmitted through the elements of our system provided a spectral bandwidth of $50\%$.  The VH is a holographic representation of the interference pattern between a vortex and a plane wave \cite{Bazhenov1990,Heckenberg1992}.  An $m = 2$ VH [see Fig. 3] provides the desired transmission function $t_2=\exp(i2\theta')$ in the first diffracted order.  Much like conventional gratings, the beam emergence angle is wavelength dependent.  Thus, a complementary diffraction grating (DG) may be used to compensate for chromatic dispersion while maintaining the vortex phase \cite{Leach2003,Mariyenko2005}.  The broadband high-contrast performance of the holographic approach compares well with other high-precision vortex phase elements \cite{Kanburapa2012conf,Kanburapa2012thesis,Errmann2013}.

\begin{figure}[t!]
\includegraphics[width=\columnwidth]{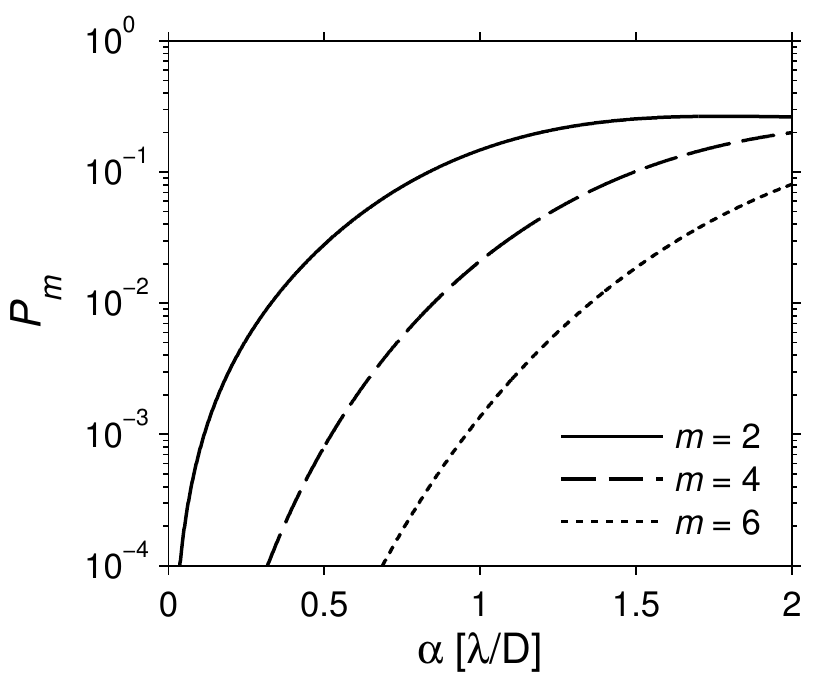}
\caption{Normalized power transmitted through the Lyot stop ($R_{LS}=R/2$) due to a distant point source displaced from the optical axis by angle $\alpha$ in units of the resolution angle $\lambda/D$.}
\end{figure}
\begin{figure}[b!]
\centerline{\includegraphics[width=\columnwidth]{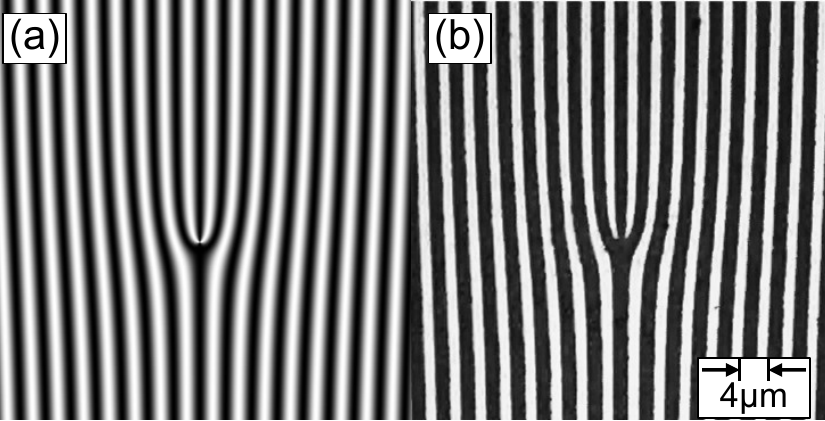}}
\caption{(a) Theoretical interference pattern between a plane wave and a charge 
$m = 2$ vortex.  (b) Zoomed-in image of the $10\unit{mm}\times10\unit{mm}$ binary amplitude grating fabricated by laser lithography with 250 line pairs per mm ($4\unit{\mu m}$ pitch).  }
\end{figure}
\begin{figure*}[htbp]
\centerline{\includegraphics[width=2.0\columnwidth]{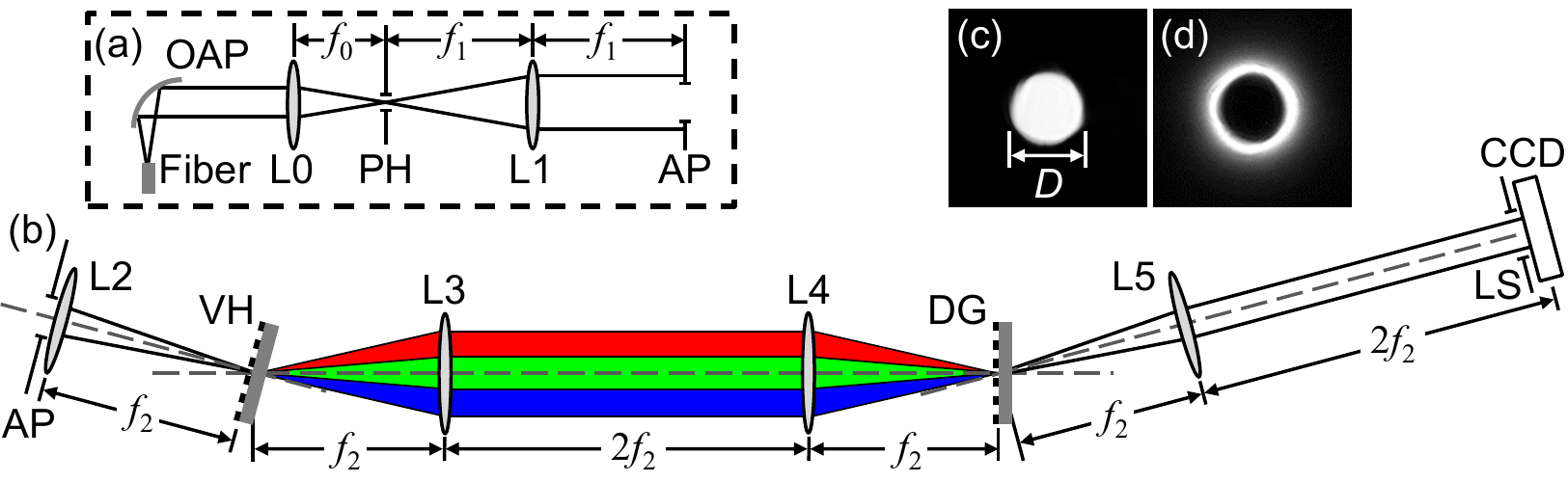}}
\caption{Experimental implementation of a broadband holographic OVC.  (a)  White light is launched from a fiber, collimated by an off-axis parabolic mirror OAP, and focused by lens L0 ($f_0 = 30\unit{mm}$) on to a small pinhole (PH).  Lens L1 ($f_1 = 30\unit{mm}$ or $100\unit{mm}$) forms the Fourier transform of PH at aperture AP ($D = 0.5\unit{mm}$) to simulate a distant point source.  Quasi-plane waves enter the OVC at AP.  (b)  Light that transmits through AP is focused by lens L2 onto vortex hologram VH [see Fig. 3(b)], which impresses an $m = 2$ vortex phase profile on the first diffracted order.  Lenses L3 and L4 reimage the focal spot at diffraction grating DG.  The line pair spacing of DG is equal to that of the VH ($4\unit{\mu m}$).  Hence, light that emerges in the first diffracted order from DG is dispersion compensated.  Lens L5 forms the exit pupil at the CCD.  Lenses L2, L3, L4, and L5 have focal length $f_2 = 300\unit{mm}$.  A variable Lyot Stop (LS) may be placed in front of the CCD to block light that originates on the optical axis at the PH plane.  (c)-(d)  Measured irradiance at the CCD without the LS in place ($5\unit{\mu m}$ diameter pinhole and $f_1 = 30\unit{mm}$).  (c)  If the pinhole is displaced from the optical axis, an approximate image of the entrance pupil appears at the CCD.  (d)  In the case where the pinhole source coincides with the optical axis, most of the light is relocated outside of the geometric exit pupil and may be blocked by LS.}
\end{figure*}

The operation of the holographic OVC [see Fig. 4] is similar to that of the schematic in Fig. 1, but the entrance pupil plane is reimaged twice.  The VH shown in Fig. 3(b) is placed at the first focal plane of the OVC optical system [see Fig 4(b)].  The light that emerges from the VH in the first diffracted order contains the desired $m = 2$ vortex phase pattern.  In addition, the light in the first order is dispersed in the transverse plane by an angle that depends on the wavelength.  The field at the first focal plane is reimaged onto a diffraction grating (DG), whose line spacing matches that of the VH.  The second grating compensates for the dispersion introduced by the VH in the first diffracted order without modifying the vortex phase.  The result is a broadband OVC wherein the field magnitude at the final pupil plane is proportional to that of the simplified system described in Fig. 1. 

To demonstrate the optical system illustrated in Fig. 4 in the laboratory, white light ($400\unit{nm}$ to $700\unit{nm}$) from a fiber-coupled plasma source (Energetiq EQ-99FC) is collimated by an off-axis parabolic mirror OAP and is subsequently focused onto a small pinhole by lens L0 ($f/1.1$, $f_0 = 30\unit{mm}$).  The pinholes used vary from $5\unit{\mu m}$ to $25\unit{\mu m}$.  Lens L1 ($f/1.1$ or $f/3.9$, $f_1 = 30\unit{mm}$ or $100\unit{mm}$) collimates the transmitted light such that the pinhole simulates an unresolved source.  The OVC entrance pupil is formed by aperture AP with diameter $D = 0.5\unit{mm}$.  AP is immediately followed by focusing lens L2 ($f/11.8$, $f_2 = 300\unit{mm}$).  An evenly illuminated entrance pupil forms an Airy disk that is centered on the central dislocation of a VH with $4\unit{\mu m}$ pitch.  The VH fabricated for this demonstration approximates the forked interference pattern with a $10\unit{mm}\times10\unit{mm}$ binary amplitude grating printed by a laser lithographic technique at $0.6\unit{\mu m}$ resolution on a $2\unit{mm}$ thick fused silica plate [see Fig 3(b)].  We approach a point phase singularity by choosing a small aperture and long focal length $f_2$ (effectively $f/600$).  This ensures the central feature is much smaller than the focal spot size.  The central lobe of the Airy disk covers approximately $256$ and $146$ line pairs for the red and blue bands, respectively.  The first diffracted order, containing an $m = 2$ vortex, passes through lens L3 ($f/5.9$, $f_2 = 300\unit{mm}$) to form a laterally dispersed pattern at the first pupil plane, where the field is imaged onto a DG with $4\unit{\mu m}$ pitch by lens L4 ($f/5.9$, $f_2 = 300\unit{mm}$).  The recombined beam passes through field lens L5 ($f/11.8$, $f_2 = 300\unit{mm}$) and forms the final pupil plane at the CCD with a central detected wavelength of 550nm.  The image of the entrance aperture appears if the pinhole is displaced from the optical axis [see Fig. 4(c)].  However, if the center of the pinhole is aligned with the optical axis, most of the light is relocated outside of the geometric image of AP [see Fig. 4(d)].  The transmission of the OVC optics (from AP to the LS) is $T=0.56\%$.  This could be enhanced by use of a blazed or volume hologram and grating.

We define the relative suppression of an extended source as $\eta_m=\kappa_m/\kappa_0$, where 
\begin{equation}
\kappa _m=\int_{0}^{2\pi }\int_{0}^{R/2}I_m\left( r'',\theta'' \right)r''dr''d\theta'',
\end{equation}
and $I_m\left( r'',\theta '' \right)$ is the irradiance at the CCD.  We note that the suppression is defined relative to the $m = 0$ case and was calculated from the signal present in the central region of the CCD image with the background signal from scattering removed.  The background is taken to be a constant, evaluated by averaging the counts in a dark region far from the optical axis (i.e. in an annulus with inner and outer diameter $15R$ and $17R$).  The interior power $\kappa_m$ is integrated over the inner half of the geometric exit pupil radius to avoid contributions from optical aberration (i.e. $R_{LS}=R/2$).  The value of $\kappa_0$ is obtained by measuring the power with the source substantially displaced from the optical axis.  We measure a maximum suppression of $\eta_2=(1.9\pm1.2)\times10^{-4}$ with an on-axis pinhole source whose angular extent is $(0.046\pm0.009)\,\lambda/D$ over a $\Delta\lambda/\lambda = 50\%$ passband ($300\unit{nm}$ band centered at $550\unit{nm}$).  The nonzero value of $\eta_2$ is partly attributed to the extended size of the pinhole source.  

\section{Sub-resolution measurements}
In cases where the structure of an unresolved source distribution is difficult to discern by conventional imaging, a non-imaging vortex-phase filtering approach may be employed to extract spatial information.  An ideal OVC removes the zero spatial frequency component of a field distribution with constant irradiance across the aperture.  In other words, the OVC completely cancels a distant on-axis point source.  Meanwhile, point sources that are minutely displaced from the optical axis are only partially suppressed.  Since small differences in source position yield a large transmission response, the OVC is sensitive to extended, yet unresolvable, source distributions \cite{Guyon2006,Mawet2010}.  In the scheme employed here, the centroid and angular extent of a spatially incoherent distribution are determined.  We demonstrate the $m = 2$ case; however, other nonzero even values of $m$ may be used in principle to make similar measurements.

\begin{figure}[b!]
\centerline{\includegraphics[width=\columnwidth]{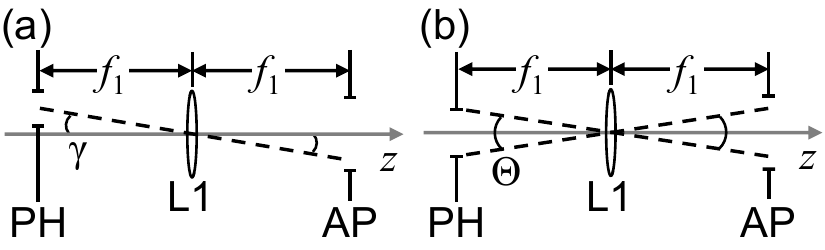}}
\caption{Diagram of pinhole source PH [see Fig. 4(a)] whose centroid is (a) misaligned and (b) aligned with the optical axis.  The angles $\gamma$ and $\Theta$ refer to the angular displacement of the source centroid and angular extent of the pinhole source, respectively.  Lens L1 forms the Fourier transform of the source at AP.}
\end{figure}

\subsection{Locating the centroid}
The centroid of an unresolved source distribution is accurately aligned with the optical axis by locating the source position that corresponds to the minimum value of $\eta_m$.  Using the experimental arrangement described in Fig. 4, we measured $\eta_2$ in the laboratory for pinhole sources with various effective angular extents $\Theta$ as a function of the angular position of the source centroid $\gamma$ [see Fig. 5].  For experimental convenience, we chose to displace the VH transverse to the optical axis rather than the pinhole [see Fig. 4(b)].  Shifting the center of the VH with respect to the focal spot is theoretically equivalent to displacing the pinhole from the optical axis.  The values of $\eta_2$ are calculated from the CCD images at the exit pupil with the background signal from scattering removed.  The post-processing of the CCD images is the dominant contribution to the experimental uncertainty.  The angular extent of the source $\Theta$ was set by varying both the diameter of the pinhole $D_{PH}$ and the value of $f_1$.  Fig. 6 shows the results for three representative sources.  The results compare well with the expected values (see e.g. \cite{Guyon2006}) and verify that the system has a steep power response to small angular displacements below the diffraction limit.  The centroid of the pinhole sources was located within $\pm0.015\,\lambda/D$, which is limited by the alignment accuracy in the laboratory.  In the case where the centroid is aligned with the optical axis (i.e. $\gamma = 0$), a nonzero signal is indicative of the angular extent of the source, as described below \cite{Swartzlander2011}. 

\begin{figure}[t!]
\includegraphics[width=\columnwidth]{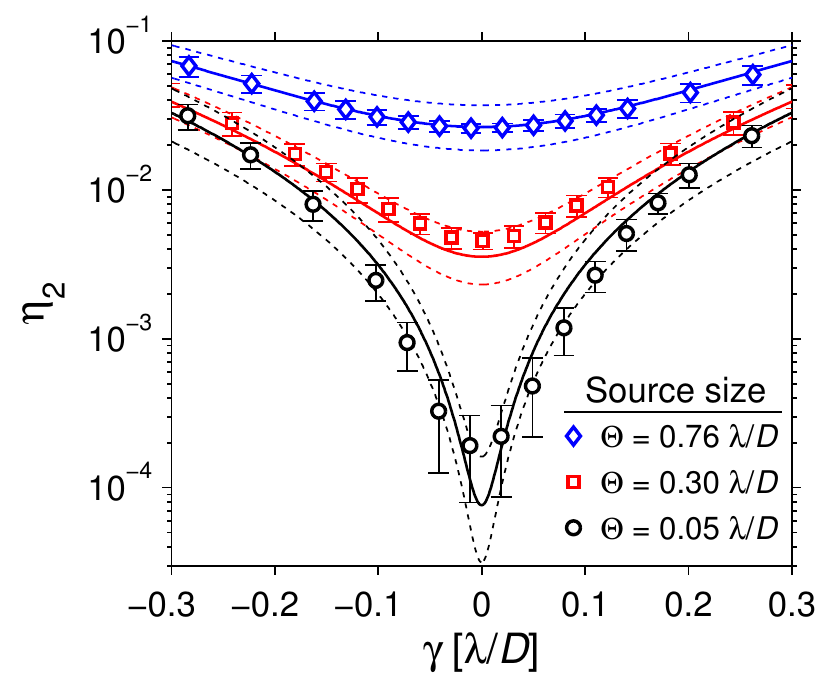}
\caption{Relative power in the Lyot Stop region of the CCD image due to three representative circular sources with angular extent $\Theta$ and angular displacement $\gamma$.  The theoretical results (solid lines) and uncertainty in the effective source size (dotted lines) are included for comparison.  The experimental error bars result from post-processing.  The power is normalized to the $m = 0$ case and a central wavelength of $\lambda = 550\unit{nm}$ is assumed.}
\end{figure}

\subsection{Determining spatial extent}
If the centroid of the source is aligned with the optical axis, the effective angular extent of the source distribution may be deduced from the value of $\eta_m$.  In the laboratory, we measure $\eta_2$ for ten values of $\Theta$ (five pinhole diameters: $D_{PH} = 5, 10, 15, 20, 25\unit{\mu m}$ and two lens L1 focal lengths: $f_1 = 30\unit{mm}$, $100\unit{mm}$).  The experimental results, shown in Fig. 7, follow the expected trend obtained numerically, which in practice may serve as a ``look up table" for the estimated source size.  The measured values for $\Theta$ have a root-mean-square deviation of $0.026\,\lambda/D$.  Discrepancies are due to misalignment and aberrations in the optical system.

In the limiting case where $\Theta$ approaches zero with respect to the angular resolution of the system, the values of $\eta_m$ have a more familiar meaning.  Assuming a localized spatially incoherent source, 
\begin{equation}
{{\kappa }_{m}}=\int_{0}^{2\pi }{\int_{0}^{{{\alpha }_{\max }}}{\mathcal{I}\left( \alpha ,\phi  \right){{P}_{m}}\left( \alpha  \right)\sin\alpha\;d\alpha\;d\phi}},
\end{equation}
where $\phi$ is the azimuthal angle about the $z$ axis and $\mathcal{I}(\alpha,\phi)$ is the source intensity distribution.  The power response due to extremely small displacements from the optical axis (i.e. $\alpha \ll \lambda/D$) is ${{P}_{m}}\propto {{\alpha }^{\left| m \right|}}$ \cite{Jenkins2008}.  Thus, the expression in Eq. (3) simplifies to the radial moments of the source distribution
\begin{equation}
{{\kappa }_{m}}=\int_{0}^{2\pi }{\int_{0}^{{{\alpha }_{\max }}}{\mathcal{I}\left( \alpha ,\phi  \right){\alpha }^{\left| m \right|}\sin\alpha\;d\alpha\;d\phi}}.
\end{equation}

Future research may examine the potential to extract spatial information about higher order moments by use of larger even values of $m$ \cite{Han2012thesis}.  Predicted differences between coherent and incoherent sources \cite{Han2012thesis} may also be explored.  The effects of systematic limitations such as vortex mask fabrication errors, lens and wavefront aberrations, as well as non-paraxial effects should also be addressed.  Further work may be directed toward achieving state-of-the-art centroid determination \cite{Lazorenko2009,Malbet2012}.

\begin{figure}[t!]
\includegraphics[width=\columnwidth]{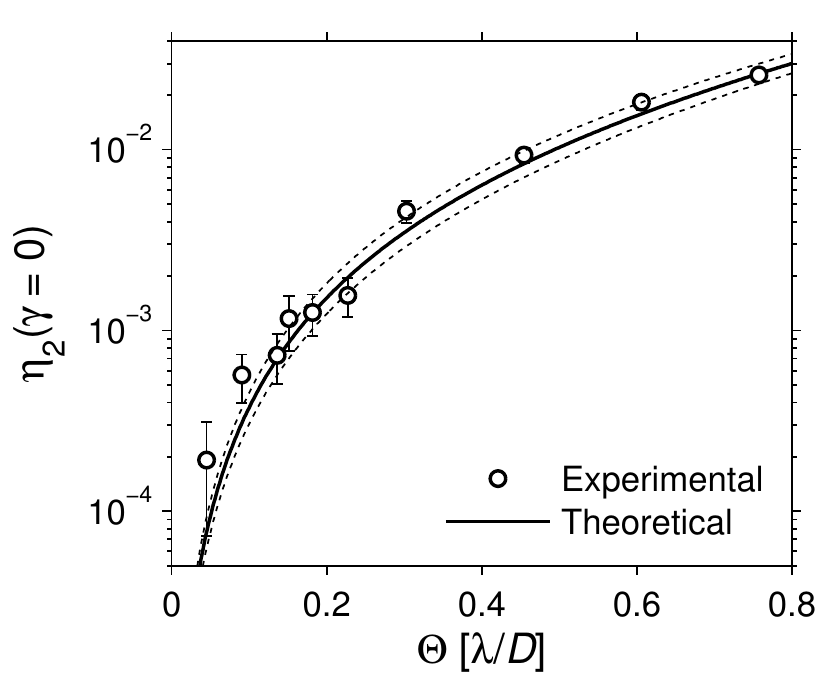}
\caption{Relative power in the Lyot Stop region of the CCD for ten different effective angular extents $\Theta$ with the source centroid located on the optical axis (i.e. $\gamma$ = 0).  The theoretical result (solid line) is shown for comparison.  The expected response owing to a $\pm10\%$ change in LS radius (dotted lines) is also included for reference.  The experimental error bars result from post-processing.  The power is normalized to the $m = 0$ case and a central wavelength of $\lambda = 550\unit{nm}$ is assumed.}
\end{figure}
\begin{figure}[b!]
\includegraphics[width=\columnwidth]{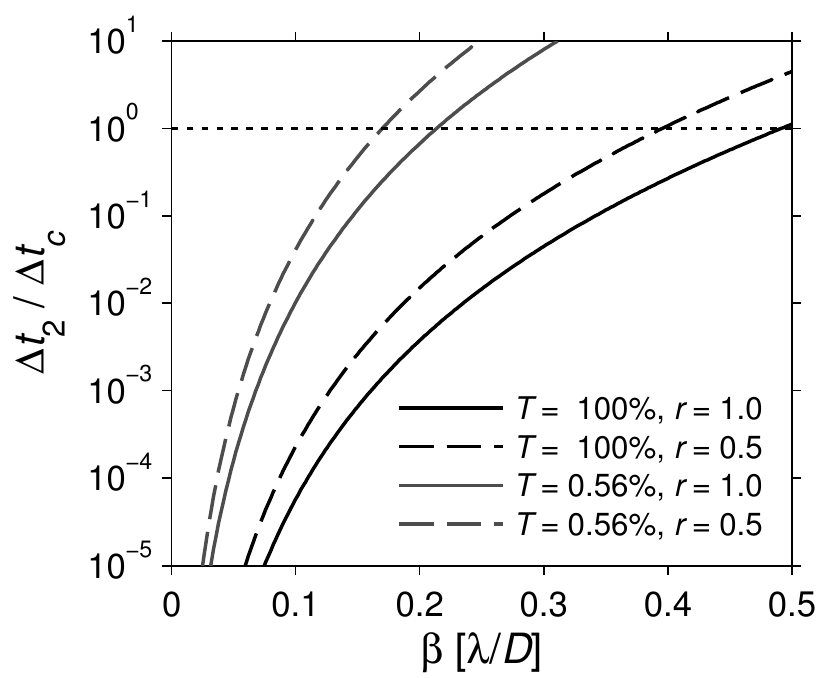}
\caption{The relative exposure time $\Delta t_2$/$\Delta t_c$ required to detect the presence of two point source separated by angle $\beta$ with a detection rate of $0.99$ and false alarm rate of $10^{-6}$.  The OVC system is compared to a conventional diffraction-limited imaging system with $N=20$ samples of the measured signal as described in \cite{Shahram2003}.  The results are plotted for an OVC with optical transmission $T$ and Lyot stop ratio $r=R_{LS}/R$.}
\end{figure}

\subsection{Noise considerations}
A potential advantage of the vortex filtering approach to sub-resolution detection is that shorter exposure times may be required relative to conventional imaging systems.  As an example, we compare the ability of each system to differentiate between two point sources of equal brightness.  In other words, we wish to determine whether a source is made up of either one or two point sources.  Assuming the OVC system is shot noise limited, the signal-to-noise ratio is proportional to the square root of the average signal measured for a specific exposure time $\Delta t_m$.  With the optical axis of the OVC centered on the combined centroid of the point sources, the exposure time required to achieve a given signal-to-noise ratio, $SNR_m$, scales as $\Delta t_m\propto (SNR_m)^2/\eta_m\kappa_0$.  The signal-to-noise for a detection rate of $0.99$ and false alarm rate of $10^{-6}$ is approximately $SNR_m=7.08$, where the shot noise distribution is taken to approach a Gaussian for a large number of photon counts.  The exposure time required for detection using an OVC relative to an equivalent conventional diffraction-limited imaging system is given by
\begin{equation}
\frac{\Delta t_m}{\Delta t_c}= \frac{1}{r^2TN^2\eta_m}\left(\frac{SNR_m}{SNR_c}\right)^2,
\end{equation}
where $\Delta t_c$ and $SNR_c$ are respectively the exposure time and signal-to-noise ratio per sample required for detection using an equivalent conventional imaging system, $T$ is the transmission of the OVC optics, $r$ is the relative Lyot stop size (i.e. $r=R_{LS}/R$), and $N^2$ is the total number of samples of the measured signal.  By statistical analysis, the $SNR_c$ required to differentiate two point sources separated by angle $\beta$ with a detection rate of $0.99$ and false alarm rate of $10^{-6}$ is approximately
\begin{equation}
SNR_c(\beta)\simeq\frac{1}{{{N}^{2}}}\left[ \frac{6.12}{{{\beta }^{4}}}-\frac{16.38}{{{\beta }^{2}}}+16.7 \right],
\end{equation}
where $\beta$ is in units of $\lambda/D$ and a Gaussian noise distribution is assumed \cite{Shahram2003}.  Thus, the relative exposure time required may be approximated by 
\begin{equation}
\frac{\Delta t_m}{\Delta t_c}\simeq\frac{0.18 N^2 \beta ^8 }{\eta_m r^2 T\left(\beta ^4-0.98 \beta
   ^2+0.37\right)^2 }.
\end{equation}
Fig. 8 shows the relative exposure time required for detecting the source separation using an $m = 2$ OVC system.  The vortex filtering approach requires substantially lower exposure times for small values of $\beta$.  We calculate that an ideal OVC ($T = 100\%$, $r = 1.0$) offers improvement for $\beta\lesssim0.49\,\lambda/D$.  For the system demonstrated above ($T = 0.56\%$, $r = 0.5$) the expected improvement is limited to $\beta\lesssim0.17\,\lambda/D$.  For cases where $\beta\ll\lambda/D$, the OVC is expected to require significantly shorter exposure times as compared to an equivalent conventional imaging system.  In practice, further improvements are possible since a single pixel photo detector may mitigate other noise contributions of CCD or CMOS sensors, such as read noise and dark current.

\section{Conclusion}
A vortex-phase filtering technique for sub-resolution information extraction has been demonstrated.  This approach allows for precision pinpointing of the centroid of an unresolved source and yields quantitative measurements of the angular extent with white light ($\Delta\lambda/\lambda=50\%$).  Spatial information is deduced from simple pupil plane power measurements at the exit pupil of a vortex coronagraph.  Here we show the angular extent $\Theta$ of an unresolved circular source may be measured for $\Theta=0.05\,\lambda/D-0.8\,\lambda/D$.  In principle, prior knowledge of the source structure is not required.  What is more, sub-resolution detection may be performed with considerably lower exposure times than conventional imaging.

\begin{acknowledgments}
We thank Mr. Thomas Grimsley, Dr. Alan Raisanen, and Dr. Stefan Preble from RIT for discussion regarding fabrication processes.  We acknowledge the Cornell Nanofabrication Facility for fabricating the holograms used in this study.  This work was supported by the U.S. Army Research Office under grant number W911NF1110333-60577PH. 
\end{acknowledgments}

\end{document}